# Un protocole de routage à basse consommation d'énergie selon l'approche de clustering pour les réseaux de capteurs sans fils

Manel Khelifi, Assia Djabelkhir

UAMB, École Doctorale en Informatique
ReSyD
Bejaïa, Algérie.
*manel.khelifi@gmail.com, assia.djabelkhir@gmail.com*

***Résumé*— Un réseau de capteur est composé d'un grand nombre de capteurs déployés dans des régions à surveiller, et communiquant entre eux à travers un médium sans fil. Le routage de données collectées dans le réseau consomme la plus grande partie d'énergie des nœuds capteurs. A cet effet, plusieurs approches de routage ont été proposées pour conserver la ressource énergétique au niveau des capteurs et pouvoir surmonter les défis inhérents à sa limitation.**

**Dans ce papier nous proposons un nouveau protocole de routage à basse consommation d'énergie pour les réseaux de capteurs basé sur l'approche de clustering. Notre protocole vise à exploiter plus équitablement l'énergie des nœuds sélectionnés cluster-heads, et à économiser l'énergie dissipée lors de l'acheminement des données capturées à la station de base.**
**Les résultats des simulations ont montré que notre protocole permet une réduction de la dissipation d'énergie et une durée de vie du réseau plus grande.**

## I. INTRODUCTION

Les réseaux de capteurs sans fil (Wireless Sensor Networks; WSN) sont considérés comme un type spécial de réseaux ad hoc, composés d'un grand nombre de capteurs matériellement petits, et placés généralement prés des objets auxquels ils s'intéressent dans les environnements où ils sont déployés. Ces capteurs sont capables de récolter, traiter et acheminer les données environnementales de la région surveillée d'une manière autonome, vers des stations de collecte appelées nœuds puits ou stations de base [1, 2].
Dans les réseaux de capteurs, la consommation d'énergie est une contrainte très cruciale puisque généralement les capteurs sont déployés dans des zones inaccessibles. Ainsi, il est difficile voire impossible de remplacer les batteries après leur épuisement. De ce fait, la consommation d'énergie au niveau des capteurs a une grande influence sur la durée de vie du réseau en entier. Il est donc impératif de mettre en place des protocole de routage efficaces en énergie, et qui prennent en compte les contraintes imposées par ces capteurs. La majorité des travaux de recherche menés actuellement se concentrent principalement sur les moyens de réduire au minimum l'énergie consommée dans la communication de données de sorte à maximiser la durée de vie du réseau.

Dans ce papier, nous proposons une nouvelle approche de routage efficace en énergie, en adaptant des techniques de clustering pour l'élaborer. Le clustering consiste à partitionner le réseau en clusters. Dans chaque cluster, des cluster-heads (CHs) sont désignés, soit par élection par les autres nœuds capteurs [3, 4], sinon ils sont assignés par une autorité centralisée [5]; nœud puits par exemple. Les CHs sélectionnés assurent non seulement la gestion de leurs clusters et la collecte de données mais aussi la transmission de ces données collectées à la station de base.

Le reste du papier est structuré comme suit: la section II, décrit notre protocole. Dans la section III, les résultats de simulation sont interprétés. Nous dressons la conclusion dans la section IV.

## II. LE PROTOCOLE PROPOSÉ

### A. Motivation

La majorité des protocoles de routage conçus pour les réseaux de capteurs de petite ou moyenne taille fournissent de bonnes performances. Cependant, lorsque le nombre de nœuds augmente, le trafic de contrôle prédomine les communications réelles. Ce qui conduit à l'augmentation dans la latence et à l'explosion des tables de routage. Afin de pallier à ces limites, les protocoles de routage pour réseaux de capteurs à topologie hiérarchique ont été introduits. Ces protocoles permettent la réduction du nombre de messages transmis, en allégeant le trafic circulant dans le réseau, réduisant ainsi la consommation d'énergie [6, 7]. Par ailleurs, la transmission de données à la station de base via un seul saut devient impossible quand l'étendue du réseau augmente. Pour remédier à ce problème, le routage à plusieurs sauts est le mode de communication adopté pour transmettre les données à la station de base [8].

### B. Principe Et Fonctionnement

Notre protocole permet d'organiser le réseau en couches afin d'assurer un routage multi-sauts entre les nœuds. Dans cet objectif nous avons défini un nouveau mécanisme de

## III. CONCLUSION

Dans ce papier, nous avons présenté un nouveau protocole de routage selon l'approche de clustering pour les réseaux de capteurs sans fil. Le protocole proposé adopte des mécanismes appropriés pour le routage à basse consommation d'énergie à travers l'introduction d'une structure en couches pour la topologie du réseau, ce qui offre une souplesse dans la communisation multi sauts des données à la station de base. Par ailleurs, notre protocole assure un équilibrage dans la charge des CHs en termes de nombre de capteurs à gérer et paquets à router dans le réseau, en apportant un partitionnement de tailles inégales en nombre de capteurs par cluster.

Nous avons démontré via les résultats de simulations que notre protocole présente de meilleures performances en terme de fiabilité, durée de vie et conservation d'énergie. Comme future travail, nous envisageons l'étude de notre protocole, dans le cas d'un réseau avec des nœuds capteurs hétérogènes et en présence de mobilité.